Extreme Spectral Risk Measures: An Application to Futures Clearinghouse Margin

Requirements

By

John Cotter and Kevin Dowd[*]

Abstract

This paper applies the Extreme-Value (EV) Generalised Pareto distribution to the extreme tails of the return distributions for the S&P500, FT100, DAX, Hang Seng, and Nikkei225 futures contracts. It then uses tail estimators from these contracts to estimate spectral risk measures, which are coherent risk measures that reflect a user's risk-aversion function. It compares these to VaR and Expected Shortfall (ES) risk measures, and compares the precision of their estimators. It also discusses the usefulness of these risk measures in the context of clearinghouses setting initial margin requirements, and compares these to the SPAN measures typically used.

Revised, December 14, 2005

[*] John Cotter is at the Centre for Financial Markets, Smurfit School of Business, University College Dublin, Carysfort Avenue, Blackrock, Co. Dublin, Ireland; email: john.cotter@ucd.ie. Kevin Dowd is at the Centre for Risk and Insurance Studies, Nottingham University Business School, Jubilee Campus, Nottingham NG8 1BB, UK; email: Kevin.Dowd@nottingham.ac.uk. Cotter's contribution to the study has been supported by a University College Dublin Faculty of Commerce research grant. Dowd's contribution was supported by an Economic and Social Research Council research fellowship on 'Risk measurement in financial institutions' (RES-000-27-0014). The authors thank two referees for very helpful comments, but the usual caveat applies.

# 1 INTRODUCTION

One of most important functions of a futures clearinghouse is to act as counterparty to all trades that take place within its exchanges. This ensures that individual traders do not have to concern themselves with credit risk exposures to other traders, because the clearinghouse assumes all such risk itself. However, it also means that the clearinghouse has to manage this risk, and perhaps the most important way it can do so is by setting margin requirements to protect itself against possible default by investors who suffer heavy losses. But how should clearinghouses set their margin requirements?

A good starting point is to assume that investor defaults are due to large – that is to say, extreme – price movements that are best analysed using some form of Extreme-Value (EV) theory, and a number of papers have followed this line of inquiry (e.g., Broussard (2001), Longin (1999, 2000), and Booth *et al* (1997)).[1] Typically, extremes are modelled by applying the unconditional Generalized Pareto Distribution (GPD) to exceedances $X$ over a high threshold $u$. The application of the GPD is justified by theory that tells us that exceedances should asymptotically follow a GPD as the threshold gets bigger. Once the GDP curve is fitted to the data, we can then extrapolate to give us estimates of any quantiles or tail probabilities we choose. Where we are interested in the extreme tails of the distribution, the GPD is far superior to alternatives such as a normal (Gaussian) distribution, which tends to under-estimate the heaviness of futures tail returns and is, in any case, inconsistent with any of the distributions that EV theory tells us to expect. Basing margin requirements on a GPD is also better than relying on historical distributions that are unable to provide very low probability estimates due to insufficient data.

These previous studies have focused on the estimation of the VaR risk measure at high confidence levels. However, the VaR has been heavily criticised as a risk measure on the grounds that it does not satisfy the properties of coherence and, in particular, because it is not subadditive (Artzner *et al.* 1999; Acerbi, 2004). The failure of VaR to be subadditive can then lead to strange and undesirable outcomes: in the present case, the use of the VaR to set margin requirements takes no account of the magnitude of possible losses exceeding VaR, and can therefore leave the clearinghouse exposed to very high losses.

---

[1] Alternative statistical modelling includes the normal distribution (e.g., Figlewski (1984)) and using the historical distribution (e.g., Warshawsky (1989)). However, these are often problematic in an EV context.

This paper suggests that it would be better to set margins on the basis of more 'respectable' risk measures. One such measure is the Expected Shortfall (ES). This is the average of the worst $100(1-\alpha)\%$ of losses, where $\alpha$ is the confidence level.[2] Unlike the VaR, the ES is coherent (and hence subadditive as well) and so satisfies many of the properties we would desire from a 'respectable' risk measure.[3] The ES also takes account of the magnitude of losses exceeding the VaR. These attractions suggest that the ES would provide a better basis for setting margin requirements than the VaR.

Alternatively, margins might be based on Spectral Risk Measures (SRMs). These risk measures have recently been proposed by Acerbi (2002, 2004), and their distinctive feature is that they relate the risk measure directly to the user's risk-aversion function. 'Well-behaved' SRMs belong to the family of coherent risk measures and therefore have the attractive properties of such measures; they are also more attractive than the ES in that they take account of user risk aversion (which the ES does not). In this paper, we consider SRMs based on an exponential risk aversion function that models the user's risk aversion in terms of a single parameter, the coefficient of absolute risk aversion. The user is the clearinghouse itself, and the coefficient of absolute risk aversion would be chosen by the clearinghouse to reflect its corporate 'attitude' towards risk. This type of SRM is therefore contingent on a single parameter whose value would in principle be straightforward to ascertain. By contrast, the VaR and the ES are contingent on a parameter, the confidence level, whose 'best' value is not easy to determine.

---

[2] This measure is closely related, but not identical to, the Tail Conditional Expectation (TCE), which is the probability-weighted average of losses exceeding VaR. For more on these risk measures and their distinguishing features, see Acerbi and Tasche (2001) or Acerbi (2004). We don't consider the TCE further in this paper because it is equivalent to the ES where the density function is continuous, and where it differs from the ES, it is not coherent.

[3] Let $X$ and $Y$ represent any two portfolios and let $\rho(.)$ be a measure of risk over a given forecast horizon. The risk measure $\rho(.)$ is subadditive if it satisfies $\rho(X+Y) \leq \rho(X) + \rho(Y)$. Subadditivity is the most important criterion we would expect a 'respectable' risk measure to satisfy, and it can be demonstrated that VaR is not subadditive unless we impose the empirically implausible requirement that returns are elliptically distributed. This makes it very difficult to regard the VaR as a 'respectable' measure of risk.

There are of course different types of margin requirement, and this paper focuses on the relatively simple problem of setting initial margins. In this problem, the clearinghouse tries to protect itself against the risk of large adverse price movements that might occur over a long period of time (see Longin (2000) for a discussion). This use of EV theory is somewhat akin to stress-testing extreme events, and implies that the clearinghouse is interested in the *unconditional* distribution of futures price movements. However, an unconditional modelling approach is less suited to the setting of daily or maintenance margins. For these latter problems, it is important that margins reflect changes in current market conditions, especially changes in market volatilities, and this requires modelling the *conditional* distribution of futures price movements. For modelling these types of margins, we would therefore need models that allow for time-varying risk measures, and these might be based on GARCH or similar volatility processes (see, for example, Barone-Adesi *et al.* (1999), McNeil and Frey (2000), Cotter (2001) or Giannopoulos and Tunaru (2004)). Such models would cater for heteroskedasticity in financial time series and lead to regular updating of futures margins (e.g., on a daily basis). However, such conditional modelling is more involved, because it would force us to address temporal dependence issues as well as the EV and financial risk measure issues that are the main focus of this paper. We therefore specify an unconditional modelling approach and limit ourselves to the setting of initial margins, and leave conditional modelling and other types of margins to a later paper.[4]

More specifically, this paper reports estimates of VaR, ES and SRMs, estimated on long and short positions in each of the S&P500, the FTSE100, the DAX, the Hang Seng, and the Nikkei225 indexes. It compares these different estimates to each other, and also examines and compares alternative methods of estimating their precision. In addition, it compares these risk measures with the types of risk measure actually used by clearinghouses for setting their initial margins, and suggests that ES and SRM risk measures would be more suitable.[5]

---

[4] An interesting alternative is to infer the initial margin requirements from risk-neutral distributions (RND) and/or implied real distributions that allow one to infer risk aversion from market data (see, for example Ait-Sahalia and Lo (2000), Bliss and Panigirtzoglou (2004), Panigirtzoglou and Skiadopoulos (2004) and Markose and Alentorn (2005)). These approaches can be used to generate forward-looking risk measures that appear to perform well. However, they yield time-varying distributions and are therefore better suited to conditional modelling problems. More importantly, such approaches yield the risk-aversion of a representative investor, whereas the risk-aversion we are interested in here is that of the clearinghouse itself.

The main objectives of this paper are therefore as follows. (1) It suggests that the risks of extreme futures positions be measured using extreme-value ES and SRM risk measures, and suggests that these might be used by futures clearinghouses to set initial margin requirements. (2) It provides estimates of these measures and of the extreme-value VaR, and also provides and compares estimates of the precision of these three risk measure estimators. (3) It discusses the suitability of these risk measures for setting margin requirements, and compares these with the measures currently used for this purpose.

This paper is organised as follows. Section 2 reviews the risk measures to be examined. Section 3 then reviews the extreme-value (EV) theory to be applied: the Peaks-Over-Threshold (POT) theory based on the Generalised Pareto distribution (GPD) applied to exceedances over a high threshold. Section 4 introduces the data and provides some preliminary data analysis on both long and short positions in five representative futures contracts. Section 5 discusses the bootstrap algorithm used. Section 6 then estimates VaR and ES, and section 7 estimates the SRMs. Each of these sections also examines the precision of estimators of these risk measures. Section 8 discusses our results and compares the suitability of each type of risk measure for futures clearinghouse margin requirements. This section also compares these risk measures with the types of risk measures actually used by clearinghouses for setting their initial margins. Section 9 concludes.

## 2. MEASURES OF RISK

Let $X$ be a random loss variable (which gives losses a positive sign and profits a negative one) over a daily horizon period on a futures contract position (which might itself be long or short in the underlying index). The VaR at the confidence level $\alpha$ is:

---

[5] Our analysis is also preliminary in another sense. We consider margin requirements on specific contracts rather than margin requirements on multiple contracts traded within a clearinghouse. We therefore do not address the (difficult) issue of how clearinghouses should aggregate from margin requirements on individual contracts to margin requirements on portfolios of contracts. Such a task would require an extensive treatment of multivariate extremes, and the issues involved can be seen in Breymann *et alia* (2003) or Poon *et alia* (2003).

$$VaR_\alpha = q_\alpha \tag{1}$$

where $q_\alpha$ is the relevant quantile of the loss distribution. However, as noted already, the VaR is highly problematic as a financial risk measure.[6]

Our second risk measure is the coherent ES, which is the average of the worst $(1-\alpha)100\%$ of losses. In the case of a continuous loss distribution, the ES is given by:

$$ES_\alpha = \frac{1}{1-\alpha} \int_\alpha^1 q_p \, dp \tag{2}$$

Using an ES measure implies taking an average of quantiles in which tail quantiles have an equal weight and non-tail quantiles have a zero weight. However, the fact that the ES gives all tail losses an equal weight suggests that someone who uses this measure is risk-neutral at the margin between better and worse tail outcomes, and is inconsistent with risk-aversion.[7]

We also want a coherent risk measure that accommodates a user's risk aversion. Let us define more general risk measures $M_\phi$ that are weighted averages of the quantiles $q_p$:

$$M_\phi = \int_0^1 \phi(p) q_p \, dp \tag{3}$$

---

[6] There are also other problems. The VaR is not consistent with expected utility maximisation, except in the very unusual case where risk preferences are lexicographic (Grootveld and Hallerbach, 2004, p. 33). More insight into the limitations of VaR comes from the downside risk literature (see, e.g., Bawa (1975), Fishburn (1977)). This suggests that we can think of downside risk in terms of lower-partial moments (LPMs), which are probability-weighted deviations of returns $r$ from some below-target return $r*$: more specifically, the LPM of order $k \geq 0$ around $r*$ is equal to $E[\max(0, r*-r)^k]$, and the parameter $k$ reflects the degree of risk aversion. The user is risk-averse if $k > 1$, risk-neutral if $k = 1$, and risk-loving if $0 < k < 1$. However, we would choose the VaR as our preferred risk measure only if $k = 0$ (Grootveld and Hallerbach, 2004, p. 35). Thus, the choice of VaR as a preferred risk measure implies a strong degree of 'negative risk aversion'.

[7] This interpretation is also confirmed from the downside risk literature. From that perspective, the ES is the ideal risk measure if $k=1$, and this implies that the user is risk-neutral (Grootveld and Hallerbach, 2004, p. 36).

for some weighting function $\phi(p)$ that reflects the user's risk aversion.[8]

We now want to know the conditions that $\phi(p)$ must satisfy in order to make $M_\phi$ coherent. The answer is the class of spectral risk measures, in which $\phi(p)$ obeys the following properties:[9]

- *Non-negativity:* $\phi(p) \geq 0$ for all $p$ belong in the range [0,1].

- *Normalization*: $\int_0^1 \phi(p)dp = 1$.

- *Increasing*ness: $\phi(p_1) \leq \phi(p_2)$ for all $0 \leq p_1 \leq p_2 \leq 1$.

The first condition requires that the weights are non-negative, and the second requires that the probability-weighted weights should sum to 1. Both of these are obvious. The third condition is a direct reflection of risk-aversion, and requires that the weights attached to higher losses should usually be bigger than, or certainly never less less than, the weights attached to lower losses. Thus, if a user has a 'well-behaved' risk-aversion function, then the weights will rise smoothly, and will also rise more rapidly, the more risk-averse the user.

To obtain a spectral risk measure, the user must specify a particular form for their risk-aversion function. A plausible candidate is an exponential utility function, which can be transformed into the following exponential risk-aversion function:

---

[8] Strictly speaking, the spectral risk measures discussed here are the measures that Acerbi describes as non-singular spectral risk measures, but we ignore the difference between singular and non-singular spectral risk measures. Note, too, that the spectral risk measure (3) also includes both the VaR and the ES as special cases. The VaR implies a $\phi(p)$ function that takes the degenerate form of a Dirac delta function that gives the outcome $p=\alpha$ an infinite weight, and every other outcome a zero weight, and the ES implies a discontinuous $\phi(p)$ that takes the value 0 for profits or small losses and takes a constant value for high losses. However, these are not 'well-behaved' spectral risk measures, because they are inconsistent with (positive) risk aversion.

[9] For more on these, see Acerbi (2004, proposition 3.4). There is also a good argument that spectral measures are the only really interesting coherent risk measures. Kusuoka (2001) and Acerbi (2004, pp. 180-182) show that all coherent risk measures that satisfy the two additional properties of comonotonic additivity and law invariance are also spectral measures. The former condition is that if two random variables $X$ and $Y$ are comonotonic (i.e., always move in the same direction), then $\rho(X+Y) = \rho(X) + \rho(Y)$; comonotonic additivity is an important aspect of subadditivity, and represents the limiting case where diversification has no effect. Law-invariance is equivalent to the (in practice essential) requirement that a measure be estimable from empirical data.

$$\phi(p) = \frac{Re^{-R(1-p)}}{1-e^{-R}} \tag{4}$$

where $R \in (0, \infty)$ is the user's coefficient of absolute risk aversion.[10] This function satisfies the conditions required of a spectral risk measure, and is attractive because it depends on a single parameter, the coefficient of absolute risk aversion. A spectral risk-aversion function is illustrated in Figure 1. This shows how the weights rise with the cumulative probability $p$, and rise more rapidly for a more risk-averse user.

**Insert Figure 1 here**

It is also curious to note that the risk aversion parameter $R$ plays a role in spectral measures similar to the role played by the confidence level $\alpha$ in the VaR and ES. Moreover, if we think in loose terms of a higher confidence level as reflecting a greater concern with higher losses – which might reflect some crude sense of increasing risk-aversion – then this is comparable to a rising $R$ in our spectral risk measure. However, whereas the confidence level is arbitrary, the value of $R$ is in principle straightforward to determine.

To obtain our spectral measure $M_\phi$, we substitute $\phi(p)$ and $q_p(X)$ into $M_\phi$ to get:

$$M_\phi = \int_0^1 \frac{Re^{-R(1-p)}}{1-e^{-R}} q_p dp \tag{5}$$

(5) gives us our third risk measure, the spectral risk measure, contingent on a chosen value of $R$.

## 3. THE PEAKS OVER THRESHOLD (GENERALISED PARETO) APPROACH

---

[10] This representation is equivalent to that of Acerbi (2004, p. 178), where his risk aversion parameter $\gamma$ is equal to the inverse of our $R$. We prefer the representation in the text because it is simpler and more intuitive.

As we are particularly interested in the extreme risks faced by the clearinghouse, we model extreme returns using an Extreme Value (EV) approach, and perhaps the most suitable of these for our purposes is the Peaks over Threshold (POT) approach (see, e.g., Embrechts *et alia* (1997) for more details). As the threshold $u$ gets large (as would be the case for the thresholds relevant to clearinghouses), this tells us that the distribution of exceedances tends to a Generalized Pareto Distribution:

$$G_{\xi,\beta}(x) = \begin{cases} 1-(1+\xi x/\beta)^{-1/\xi} & \quad \xi \geq 0 \\ 1-\exp(-x/\beta) & \quad \xi < 0 \end{cases} \quad (6)$$

where

$$x \in \begin{cases} [0,\infty) \\ [0,-\beta/\xi] \end{cases} \quad \text{if} \quad \begin{array}{l} \xi \geq 0 \\ \xi < 0 \end{array}$$

and $\xi$ and $\beta > 0$ are the shape and scale parameters conditional on the threshold $u$.

Taking high quantiles representing high losses, the $p^{\text{th}}$ quantile of the return distribution – which is also the VaR at the (high) confidence level $p$ – is given by:

$$q_p = VaR_p = u + \frac{\beta}{\xi}\left\{\left(\frac{n}{N_u}p\right)^{-\xi} - 1\right\} \quad (7)$$

and the ES is given by:

$$ES_p = \frac{q_p}{1-\xi} + \frac{\beta - \xi u}{1-\xi} \quad (8)$$

To obtain more general spectral risk measures, we substitute (7) into (5) to obtain:

$$M_\phi = \int_0^1 \frac{Re^{-R(1-p)}}{1-e^{-R}}\left[u + \frac{\beta}{\xi}\left\{\left(\frac{n}{N_u}p\right)^{-\xi} - 1\right\}\right]dp \quad (9)$$

Estimates of our risk measures are then obtained by estimating/choosing the relevant parameters and plugging these into the appropriate risk measure equation (i.e., (7), (8) or (9)). This is straightforward where our risk measures are the VaR and the ES; where our risk measures are spectral, we can solve (9) using a suitable numerical integration method.[11]

## 4. PRELIMINARY DATA ANALYSIS

Our data set consists of daily geometric returns (taken as the difference between the logs of respective end-of-day prices) for the most heavily traded index futures – that is, the S&P500, FTSE100, DAX, Hang Seng and Nikkei 225 futures – between January 1 1991 and December 31 2003. The data were obtained from Datastream with the contracts trading on the CME (in the cases of the S&P500 and Nikkei 225), LIFFE (in the case of the FTSE100), EUREX (in the case of the DAX) and HKSE (in the case of the Hang Seng). These data refer to futures contracts which expire in the next trading month, and the rollover from an expiring contract to the next one occurs at the start of each trading month. Datastream deals with bank holidays by padding the dataset and taking the bank holiday's end-of-day price to be the previous trading day's end-of-day price. This means that we have the same number of daily returns (i.e., 3392) for all contracts.

As a preliminary, Figure 2 shows QQ plots for these contracts' empirical return distributions relative to a normal (or Gaussian) distribution and indicates the fat-tailed property of futures. In addition, the points where the QQ plots change shape provide us with natural estimates of tail thresholds. These lead us to select thresholds of 2% for the S&P, DAX, Hang Seng and Nikkei indices and 1.5% for the FTSE.

**Insert Figure 2 here**

We find that the tail indexes are stable and adequately modelled with GPDs.[12] Maximum likelihood estimates of the GPD parameters are given in Table 1 for both long and short trading positions. The tail indices are positive except for the Nikkei and the estimated scale parameters fluctuate around 1. All of these estimates are plausible and in line with those reported from other studies. The Table also gives the assumed thresholds $u$, the associated numbers of exceedances ($N_u$) and the observed exceedance probabilities ($prob$). The numbers and probabilities of exceedances vary somewhat, but all confirm that the chosen thresholds are in the stable tail-index regions identified in Figure 2.

---

[11] More details on such methods can be found in standard references (e.g., Miranda and Fackler (2002, chapter 5).

[12] The goodness of fit of the GPD is confirmed by plots of the tail index against the number of exceedances, and mean-excess function plots fitted to GPDs. These are available on request.

**Insert Table 1 here**

## 5. BOOTSTRAP ALGORITHM

The estimates of standard errors and confidence intervals reported in this paper were obtained using a semi-parametric bootstrap based on the ML estimates of the GPD parameters just explained.[13] In particular, we first take 5000 bootstrap resamples, each of which consists of $n$=3392 uniform random variables. For each resample, these uniform random numbers are put in ascending order so that they can be considered as a set of resampled cumulative probabilities (or confidence levels). For the $j^{\text{th}}$ resample, let us denote this set of resample cumulative probabilities as $p_1^j, p_2^j, ..., p_n^j$, where $p_i^j \leq p_{i+1}^j$ by construction. We then use the fitted GPD (i.e., (7)) to infer each element of the $j^{\text{th}}$ resample set of losses (or quantiles). Thus, if $p_i^j$ is the $i^{\text{th}}$ cumulative probability in the $j^{\text{th}}$ resample, then $q_i^j$, the $i^{\text{th}}$ highest loss in the $j^{\text{th}}$ resample (or, equivalently, the $p_i^j$ quantile of the $j^{\text{th}}$ resample), can be obtained from (7), i.e.,

$$q_i^j = \hat{u} + \frac{\hat{\beta}}{\hat{\xi}} \left\{ \left( \frac{n}{\hat{N}_u} p_i^j \right)^{-\hat{\xi}} - 1 \right\} \tag{10}$$

---

[13] The obvious alternative is to bootstrap from the distribution of sample returns, but if we are to retain the benefits of using an EV approach this would then require that we re-estimate the GPD parameters for each resample. However, some of these resample estimates would be degenerate (and this is especially a problem with estimates of the tail index) and this would necessitate some ad hoc intervention to rule out 'implausible' parameter values. The semi-parametric method used in the text avoids this problem and is simpler to implement.

where '^' are the sample-based estimates of the various GPD parameters. Since the quantiles are also VaRs, (10) also gives us resample estimates of the VaRs as well. Resample estimates of the ES and SRM are then obtained by using (8) and (5) respectively (with $q_p$ replaced by $q_i^j$ and the parameters replaced by their '^' estimates). For each resample, the precision estimates (e.g., the standard errors, confidence intervals, etc.) were obtained from the set of resample estimates of the relevant risk measures.[14]

## 6. ESTIMATION OF VAR AND EXPECTED SHORTFALL

Estimates of VaRs and ESs based on these GPD parameters in Table 1 are shown in Table 2 and illustrated in Figures 3 and 4. These risk measures are based on very high confidence levels and reflect the clearinghouses' concerns with very high trading losses and associated possibilities of investor default. The ESs are notably larger than the VaRs, but in general they behave in similar ways. Estimates of both risk measures increase as the confidence levels get bigger; are lowest for the S&P and FTSE contracts and highest for the Hang Seng; and show little difference between short and long positions.

**Insert Table 2 here**

**Insert Figure 3 here**

**Insert Figure 4 here**

We now examine the precision of the risk estimates. Table 3 presents some estimates of their standard errors based on a non-parametric bootstrap with 5000 resamples and the parameter values shown in Table 1. The results presented show that the ES standard errors are higher than the VaR standard errors for all contracts except the Nikkei 225. This would suggest that the VaR is more precisely estimated than the ES.

---

[14] To elaborate further, if we were seeking to estimate the confidence interval for a VaR, say, we would obtain 5000 resample estimates of the VaR (as explained in the text), order them in ascending order, and take the bounds of the 90% confidence interval for the VaR as given by the $5000 \times 0.05 = 250^{th}$ and $5000 \times 0.95 = 4750^{th}$ largest resample estimates. Note that these bounds involve a 'basic' bootstrap without any adjustments to deal with possible bias (e.g. as in the bootstraps set out in the early chapter of Efron and Tibshirani (1993)). Such adjustments were not employed because bias is essentially a small-sample problem and bias-correction refinements would make little difference given the large sample sizes available to us here.

**Insert Table 3**

However, if we compare Tables 2 and 3, we also see that the ESs have higher coefficients of variation than the VaRs, and this might suggest that the ESs are more precisely estimated than the VaRs. (The coefficient of variation is the estimated risk measure divided by its estimated standard error.)

To investigate further, Table 4 also presents bootstrapped estimates of the 90% confidence intervals for the two risk measures, standardized (i.e., divided) by the means of the bootstrapped estimates to facilitate comparison. These confidence intervals are narrower for the ES than for the VaR, and again suggest that the ES is estimated (relatively) more precisely than the VaR.

**Insert Table 4 here**

These results also indicate a second interesting finding. The estimated confidence intervals are more or less symmetric for low confidence levels, but for higher confidence levels (and especially a confidence level of 0.999), they are notably asymmetric, with the right bound being further away from the mean of the bootstrapped estimates than the left bound.

## 7. ESTIMATION OF SPECTRAL RISK MEASURES

As noted already, to apply a spectral risk measure we need to choose a suitable value for the coefficient of absolute risk-aversion $R$. In principle, this can be any positive value, but in the present context it only makes sense to work with high values of $R$. The reason can be seen from the $\phi(p)$ functions shown in Figure 1. The higher is $R$, the more we care about the higher losses relative to the others. It therefore only makes sense to apply an EV approach in the first place if we care a great deal about the very high losses (i.e., the extremes) relative to the non-extreme observations, and this requires that $R$ take a high value (e.g., well above 20, and possibly much higher).

Having chosen a value for $R$, we can calculate the integral (9) using numerical integration. This approximates the continuous integral by a discrete equivalent: we discretise the continuous variable $p$ into a number $N$ of discrete slices, where the approximation gets better as $N$ gets larger, and then choose a suitable numerical integration method. The ones we considered were trapezoidal and Simpson's rules, and numerical integration using pseudo-Monte Carlo and quasi-Monte Carlo methods, with the latter based on Niederreiter and Weyl algorithms.[15]

To evaluate the accuracy of these methods, Table 5 gives estimates of the approximation errors generated by these alternative numerical integration methods based on various values of $N$ and a plausible set of benchmark parameters.[16] These results indicate that all methods have a negative bias for relatively small values of $N$, but the bias disappears as $N$ gets large. They also show that the Simpson's and trapezoidal methods are a little more accurate than the quasi-methods. This conclusion is supported by the plots in Figure 5, which show how rapidly the different integration methods converge to the 'true' values as $N$ gets large.

**Insert Table 5 here**

**Insert Figure 5 here**

For the purposes of the remaining estimations, we selected a benchmark method consisting of the trapezoidal rule calibrated with $N$=1 million, and the results just examined suggest that this benchmark should deliver highly accurate estimates.

Estimates of the spectral-exponential risk measures themselves are given in Figure 6 and Table 6. The Figure shows plots of estimated spectral measures against $1-1/R$. (We can take $1-1/R$ as a surrogate for the degree of risk-aversion, but one which is particularly convenient here because $1-1/R \rightarrow 1$ as the user becomes extremely risk averse.) These plots turn out to be very similar to the VaR and ES curves considered earlier, but with $\alpha$ replaced here by $1-1/R$. The estimated SRMs are also similar to the earlier ones in that they are lowest for the S&P and FTSE, highest for the Hang Seng, and show little difference between short and long positions.[17]

---

[15] However, results for pseudo methods are not reported here because they were considerably less accurate than the other methods.

[16] These benchmark parameters were the mean long-position parameters in Table 1 combined with $R = 100$.

**Insert Figure 6 here**

**Insert Table 6 here**

Table 7 presents estimates of the standard errors of the SRMs based on a non-parametric bootstrap procedure similar to that used earlier for the VaR and ES.[18] These results show that the estimated standard errors are broadly similar across futures contracts, but increase substantially as $R$ gets larger.[19]

**Insert Table 7 here**

Table 8 presents estimates of the standardised confidence intervals for the SRMs. Again, these are similar across different positions, but also expand markedly as $R$ rises. The confidence intervals for very high $R$ values also show small asymmetries similar to those exhibited earlier by VaR and ES confidence intervals predicated on extremely high confidence levels.

**Insert Table 8 here**

---

[17] It is also interesting to note the rough magnitudes of the spectral and earlier risk measures. In particular, it turns out that the VaR at the 0.995 confidence level, the ES at the 0.99 confidence level, and the SRM with $R = 100$ are all of much the same size. From this it follows, for example, that the SRM for $R>100$ is larger than the VaR at the 0.995 confidence level, and so on.

[18] As with the earlier bootstrap, this involves resampling from the estimated distribution function. However, in doing so we also have to restrain the number of slices $N$ to the sample size: the non-parametric bootstrap therefore involves $N$=3392.

[19] In terms of a rough order of magnitude, these results suggest that a spectral risk measure with $R = 100$ usually has standard errors larger than those of the VaR or ES at the 0.995 confidence level, but less than those of the VaR or ES at the 0.999 confidence level.

However, we also see that SRMs have considerably wider confidence intervals than the VaR and ES. Thus, our results present clear evidence that estimators of SRMs are less precise than estimators of VaR or ES. The explanation appears to be related to effective sample size, and can be seen by comparing the ES and SRM estimators. If we have $n$ observations in the tail, then the ES estimator is based on a 'straight' sample of size $n$. By contrast, an SRM estimator predicated on a high value of $R$ is a weighted average of observations that puts a lot of weight on a small subset of these tail observations, and therefore operates with a smaller effective sample size.

## 8. DISCUSSION

*8.1 Comparison of the magnitude of different risk measure estimates*

All our estimated risk measures show considerable similarity. All agree that the S&P and FTSE contracts are the least risky indices, and that the Hang Seng is the most risky. The use of any of these measures for setting initial margin requirements would therefore lead to the S&P and FTSE having the lowest margin requirements and to the Hang Seng having the highest. All estimated risk measures also agree that there is only mild asymmetry across long and short positions, which suggests that there should be only small differences between the margin requirements of long and short positions. Plots of each risk measures against its conditioning parameter (i.e., the confidence level or risk aversion parameter) also show much the same exponential-like shape: this shows that the different risk measures change in much the same way when their conditioning parameter changes.

*8.2 Comparison of the precision of different risk measure estimators*

Turning now to the precision metrics, we find that the precision of all our risk measure estimators falls as the conditioning parameter rises. We also find some evidence that the confidence intervals of our risk measures tend to be become slightly asymmetric when the conditioning parameter is extremely high. In addition, we find that different precision metrics sometimes suggest slightly different pictures of the relative precision of VaR and ES estimators. For example, we might conclude that VaR is a little more precisely estimated than the ES if we use the standard error as our precision metric, but we would come to the opposite conclusion if we used the other precision metrics. However, all precision metrics also agree that estimators of SRMs are somewhat less precise than estimators of the VaR or of the ES.

### 8.3. Comparative usefulness of our risk measures for setting initial margins

We now consider the usefulness of each of these risk measures for setting initial margin requirements:

- The VaR is of limited use because it gives the clearinghouse no indication of how big its losses might be in the event that an investor suffers losses that exhaust its margin. This is the case even when the VaR is based on an extreme confidence level. Furthermore, the non-subadditivity of the VaR also creates another problem for a clearinghouse that bases margin requirements on a VaR: investors might be tempted to break up their accounts to reduce overall margin requirements, and in so doing leave the clearinghouse exposed to a hidden residual risk against which the clearinghouse has no effective collateral from its investors. This type of problem does not arise with subadditive risk measures.

- The ES is more useful because it has the benefits of coherence (and therefore subadditivity). More concretely, the ES takes account of the sizes of tail losses and has the helpful interpretation that it tells the clearinghouse the loss an investor can expect to make conditional on it experiencing a loss that exceeds a chosen threshold. The clearinghouse could set a target extreme tail probability $p$ (e.g., $p$=0.1%), and base its margin requirements on the ES at the confidence level 1-$p$. The margin requirement set this way then has a natural interpretation as the amount needed to cover the expected worst 100$p$% of loss outcomes. However, this still leaves the clearinghouse the problem of determining what the tail probability $p$ should be.

- The spectral risk measures enable us to overcome this latter problem. They are also coherent, and have the advantage that they alone take account of the user's (i.e., the clearinghouse's) degree of risk aversion. This gives clearinghouse risk managers an opportunity to select an $R$ value that reflects the clearinghouse's corporate risk aversion. Thus, spectral measures not only take account of the clearinghouse's risk aversion, but have a parameter whose value can be ascertained from it. The initial margin requirement could then be set equal to the SRM without the arbitrariness of having to specify a tail probability. The more risk-averse the clearinghouse, the greater the margin requirement. Of course, a greater margin requirement is also more costly for investors who use the exchange, but this is exactly as it should be, and it is for each clearinghouse to decide its own preferred tradeoff between safety and cost.[20]

*8.4. Comparison with the measures currently used to set initial margins*

Currently, clearinghouses set margin requirements using systems of standardized stress tests. The most common are of the SPAN type, first introduced by the Chicago Mercantile Exchange in 1988.[21] Versions of SPAN systems are now widely used by other exchanges (e.g., such as the London SPAN system used by the London Clearing House). To determine a margin requirement, the system first identifies a set of contract 'families', each family being the set of contracts sharing the same underlying asset. For each underlying, the system then estimates a spanning range and this would typically be a range covering 99% of historical one-day movements in the underlying price within the rolling data window.[22] For each contract within the family, the system produces a range of projected loss scenarios based on this set of underlying price scenarios, taking account of any relevant factors such as respective volatilities and the sensitivities of contract prices to underlying prices. The system then identifies the particular price-change scenario that produces the worst-case loss for the family as a whole, and this loss determines the margin requirement for contracts based on that underlying. The margins requirements for a portfolios of contracts across different underlyings are then obtained by adding up the individual margin requirements so obtained, with some offsets allowed to accommodate risk diversification. These offsets are determined by the clearinghouse itself, and vary from one clearinghouse to another.

---

[20] A good analysis of this tradeoff and the issues it entails is given by Shanker and Balakrishnan (2005).

[21] For more details on SPAN systems, see, e.g., Knott and Mills (2002) and LCH (2002).

[22] There is also considerable room for discretion, and a SPAN system might allow for possible 'what if?' scenarios or adjustments for market moves that not captured in the data window. Furthermore, the spanning range used would be guided by the clearinghouse's risk management policy. For example, the London SPAN system used by the LCH is guided by its core policy which is to establish levels that cover a minimum of three standard deviations of historic volatility based on the higher of one-day or two-day price movements over the previous 60 days (LCH, 2002, p. 20).

However, SPAN systems have a number of deficiencies. They require the clearinghouse to specify a tail probability for each position (which corresponds to the probability covered in the spanning range), but also require ancillary decisions on issues such as how to offset long and short positions and how to value positions that are non-linear in underlying prices.[23] But perhaps of more concern is the way in which the risks of positions in different 'families' of contracts are aggregated to produce the overall margin requirement on a diversified portfolio. In fact, the degree of offset allowed on portfolios of contracts across different underlyings is typically both limited and arbitrary. Thus SPAN systems limit the ability of investors to enjoy the benefits of diversification, and make it difficult to justify the margin requirements in statistically 'respectable' terms. The resulting risk measures can then be regarded as reflecting some implicit correlation or dependence structure that has not been explicitly set out or justified, and whose validity cannot be taken for granted. As a result, clearinghouse risk managers do not in fact know whether their SPAN model has a defensible dependence structure or not. There is also the related problem that the degree of cover it provides the clearinghouse is also unclear.

Many of these deficiencies can be overcome by using ES and SRM measures. For instance, margin systems based on ES and SRM measures would embody more explicit and therefore potentially more defensible correlation (or dependence) assumptions, and also give clearinghouses a much clearer idea of the cover that the margin requirements provide.

## 9. SUMMARY AND CONCLUSIONS

By acting as a counterparty in all trades, a clearinghouse relieves individual traders of credit risk concerns but acquires credit-risk exposures of its own. It then seeks to manage these exposures by imposing margin requirements. However, even then the clearinghouse is still exposed to the risk of loss arising from investor defaults triggered by extreme price movements. This paper has sought to model these risks using an EV approach, where initial margin requirements might be set using one of three different financial risk measures: a VaR, an ES, and an exponential SRM. Our discussion suggests the VaR should not be used as it is incoherent and takes no account of clearinghouse risk aversion. The ES is much better, both because it is coherent and because it has a natural interpretation as a margin requirement (i.e., it providing the clearinghouse with a level of cover that matches the expected worst tail losses). However, the ES has the disadvantage that it does not reflect any risk aversion on the clearinghouse's part. In contrast, SRMs have the advantages of being coherent and of taking account of clearinghouse risk aversion. They also avoid the arbitrariness of having to specify a confidence level or tail probability; instead, they require a risk-aversion parameter that can be determined from the clearinghouse's corporate 'attitude' toward risk. Thus, a clear pecking order emerges: the ES is (much) better in principle than the VaR, but the SRM is better in principle than the ES.

---

[23] Typically, spreads are allowed to be perfectly offset, but subject to an offset charge, and non-linear positions are handled using the Black (1976) model. See Knott and Mills (2002, p. 169).

However, it is also important to examine how our risk measures compare quantitatively as well. For the futures contracts considered here, we find that all three risk measures generate comparable results. For example, they all suggest that the S&P and FTSE are the least risky contracts, and that the Hang Seng is the most risky contract. They are also similar in that they all depend on a key conditioning parameter – the confidence level in the case of the VaR and ES, and the coefficient of absolute risk aversion in the case of the exponential SRMs – and increase in comparable (and intuitively plausible) ways as the conditioning parameter increases. Broadly speaking, we also find that estimators of the VaR and of the ES would appear to have fairly similar degrees of precision , but SRM estimators based on the high risk-aversion coefficients relevant in this context are somewhat less precise. Thus, SRMs would appear to be superior risk measures in principle, but have the practical disadvantage that their estimators are likely to be somewhat less precise than estimators of the older risk measures.

**REFERENCES**

Acerbi, C., Tasche, D., 2001. Expected shortfall: a natural alternative to value at risk. Economic Notes, 31, 379-388.

Acerbi, C., 2002. Spectral measures of risk: a coherent representation of subjective risk aversion. Journal of Banking and Finance, 26, 1505-1518.

Acerbi, C., 2004. Coherent representations of subjective risk-aversion. In G. Szegö (Ed.), Risk Measures for the 21$^{st}$ Century, Wiley, New York, pp. 147-207.

Ait-Sahalia, Y., A. W., 2000. Nonparametric risk management and implied risk aversion. Journal of Econometrics, 94, 9-51.

Artzner, P., Delbaen, F., Eber, J.-M., Heath, D. 1999. Coherent measures of risk. Mathematical Finance, 9, 203-228.

Barone-Adesi, G., Giannopoulos, K. and Vosper, L., 1999. VaR without correlations for portfolios of derivative securities. Journal of Futures Markets, 19, 583-602.

Bawa, V. S., 1975. Optimal rules for ordering uncertain prospects. Journal of Financial Economics, 2, 95-121.

Bliss, R. R., N. Panigirtzoglou, 2004. Option-implied risk aversion estimates. Journal of Finance, 59, 407-446.

Booth, G. G., Brousssard, J.P., Martikainen, T., Puttonen, V., 1997. Prudent margin levels in the Finnish stock index futures market. Management Science, 43, 1177-1188.

Breymann, W., A. Dias, and P. Embrechts, 2003. Dependence structures for multivariate high-frequency data in finance. *Quantitative Finance*, 3, 1-14.

Broussard, J. P. 2001. Extreme-value and margin setting with and without price limits, Quarterly Review of Economics and Finance, 41, 365-385.

Cotter, J. 2001. Margin exceedences for European stock index futures using extreme value theory. Journal of Banking and Finance, 25, 1475-1502.

Efron, B., and R. J. Tibshirani, 1993, An Introduction to the Bootstrap, (Chapman and Hall, New York).

Embrechts, P., Kluppelberg C., Mikosch, T., 1997. Modelling Extremal Events for Insurance and Finance. Springer Verlag, Berlin.

Figlewski, S. 1984. Margins and market integrity: margin setting for stock index futures and options. The Journal of Futures Markets, 4, 385–416.

Fishburn, P. C. 1977. Mean-risk analysis with risk associated with below-target returns. American Economic Review, 67, 116-126.

Giannopoulos, K., Tunaru, R., 2004. Coherent risk measures under filtered historical simulation. Journal of Banking and Finance, Forthcoming.

Grootveld, H., Hallerbach, W. G., 2004. Upgrading value-at-risk from diagnostic metric to decision variable: a wise thing to do?, in G. Szegö (Ed.) Risk Measures for the 21$^{st}$ Century. Wiley, New York, pp. 33-50.

Knott, R. and A. Mills, 2002, Modelling risk in central counterparty clearing houses: a review, Financial Stability Review, December, 162-174.

Kusuoka, S., 2001. On law invariant coherent risk measures. Advances in Mathematical Economics, 3, 83-95.

London Clearing House, 2002, Market Protection, The role of LCH: regulatory framework, structure and governance, legal and contractual obligations, risk management, default rules, financial backing. London: LCH.

Longin, F., 1999. Optimal margin levels in futures markets: extreme price movements. Journal of Futures Markets, 19, 127-152.

Longin, F., 2000. From value at risk to stress testing: The extreme value approach. Journal of Banking and Finance, 24, 1097-1130.

Markose, S., Alentorn, A. 2005. The Generalised Extreme Value (GEV) Distribution, Implied Tail Index and Option Pricing, Department of Economics Working Paper, University of Essex.

McNeil, A. J., and Frey, R., 2000. Estimation of tail-related risk measures for heteroscedastic financial time series: An extreme value approach, Journal of Empirical Finance, 7, 271-300.

Miranda, M. J., Fackler, P. L., 2002. Applied Computational Economics and Finance. MIT Press, Cambridge MA and London.

Panigirtzoglou, N., Skiadopoulos, G. 2004. A New Approach to Modeling the Dynamics of Implied Distributions: Theory and Evidence from the S&P 500 Options, Journal of Banking and Finance, 28, 1499-1520.

Poon, S.-H., Rockinger, M. and Tawn, J., 2004. Extreme-Value Dependence in Financial Markets: Diagnostics, Models and Financial Implications, Review of Financial Studies, 17, 581-610.

Shanker, L., Balakrishnan, N. 2005. "Optimal clearing margin, capital and price limits for futures clearinghouses." Journal of Banking and Finance 29: 1611-1630.

Warshawsky, M. J. 1989. The adequacy and consistency of margin requirements: the cash, futures and options segments of the equity markets. The Review of Futures Markets, 8, 420–437.

## Table 1: GPD Parameters for Futures Indexes

| Futures index | Long position | | | | | Short position | | | | |
|---|---|---|---|---|---|---|---|---|---|---|
| | $u$ | $prob$ | $N_u$ | Tail $\hat{\xi}$ | Scale $\hat{\beta}$ | $u$ | $prob$ | $N_u$ | Tail $\hat{\xi}$ | Scale $\hat{\beta}$ |
| S&P500 | 2.00 | 0.04 | 130 | 0.18 | 0.60 | 2.00 | 0.03 | 118 | 0.13 | 0.76 |
| | | | | (0.10) | (0.08) | | | | (0.15) | (0.13) |
| FTSE100 | 1.50 | 0.07 | 250 | 0.10 | 0.71 | 1.50 | 0.08 | 276 | 0.02 | 0.73 |
| | | | | (0.08) | (0.07) | | | | (0.07) | (0.07) |
| DAX | 2.00 | 0.07 | 235 | 0.01 | 1.19 | 2.00 | 0.07 | 237 | 0.05 | 1.00 |
| | | | | (0.05) | (0.10) | | | | (0.07) | (0.10) |
| Hang Seng | 2.00 | 0.10 | 353 | 0.13 | 1.18 | 2.00 | 0.11 | 367 | 0.14 | 1.15 |
| | | | | (0.06) | (0.10) | | | | (0.05) | (0.09) |
| Nikkei 225 | 2.00 | 0.08 | 277 | -0.01 | 0.89 | 2.00 | 0.08 | 255 | -0.07 | 1.04 |
| | | | | (0.06) | (0.07) | | | | (0.05) | (0.08) |

*Notes*: The Table presents estimates of the GPD parameters for long and short futures positions in the five contracts shown, where returns are expressed in daily % form. The sample size $n$ is 3392, the threshold is $u$, the probability of an observation in excess of $u$ is $prob$, the number of exceedences in excess of $u$ is $N_u$, the estimated tail parameter is $\hat{\xi}$ and the estimated scale parameter is $\hat{\beta}$. The numbers in brackets are the estimated standard errors of the parameters concerned. The thresholds $u$ are chosen as the approximate points where the QQ plots in Figure 2 change slope.

**Table 2: Estimates of GPD VaRs and Expected Shortfalls for Futures Positions**

| Futures index | Long position | | | | Short position | | | |
|---|---|---|---|---|---|---|---|---|
| | $\alpha$=0.98 | $\alpha$=0.99 | $\alpha$=0.995 | $\alpha$=0.999 | $\alpha$=0.98 | $\alpha$=0.99 | $\alpha$=0.995 | $\alpha$=0.999 |
| **VaR** | | | | | | | | |
| S&P500 | 2.414 | 2.912 | 3.476 | 5.092 | 2.436 | 3.029 | 3.677 | 5.428 |
| FTSE100 | 2.489 | 3.070 | 3.692 | 5.315 | 2.539 | 3.063 | 3.594 | 4.857 |
| DAX | 3.488 | 4.326 | 5.170 | 7.152 | 3.291 | 4.042 | 4.819 | 6.731 |
| Hang Seng | 4.171 | 5.231 | 6.392 | 9.526 | 4.190 | 5.250 | 6.419 | 9.611 |
| Nikkei 225 | 3.243 | 3.850 | 4.452 | 5.833 | 3.315 | 3.957 | 4.568 | 5.877 |
| **ES** | | | | | | | | |
| S&P500 | 3.237 | 3.844 | 4.532 | 6.503 | 3.375 | 4.056 | 4.801 | 6.813 |
| FTSE100 | 3.388 | 4.033 | 4.725 | 6.527 | 3.305 | 3.840 | 4.382 | 5.670 |
| DAX | 4.705 | 5.551 | 6.404 | 8.406 | 4.411 | 5.202 | 6.020 | 8.033 |
| Hang Seng | 5.851 | 7.070 | 8.404 | 12.007 | 5.884 | 7.117 | 8.475 | 12.188 |
| Nikkei 225 | 4.112 | 4.712 | 5.308 | 6.677 | 4.201 | 4.801 | 5.372 | 6.595 |

*Notes*: Estimates in daily % return terms based on the parameter values shown in Table 1, where $\alpha$ is the confidence level and the holding period is 1 day.

**Table 3: Standard Errors for VaRs and Expected Shortfalls**

| Futures index | Long position | | | | Short position | | | |
|---|---|---|---|---|---|---|---|---|
| | $\alpha$=0.98 | $\alpha$=0.99 | $\alpha$=0.995 | $\alpha$=0.999 | $\alpha$=0.98 | $\alpha$=0.99 | $\alpha$=0.995 | $\alpha$=0.999 |
| **VaR** | | | | | | | | |
| S&P500 | 0.0811 | 0.1311 | 0.2028 | 0.6386 | 0.0977 | 0.1500 | 0.2331 | 0.6555 |
| FTSE100 | 0.0954 | 0.1448 | 0.2195 | 0.5909 | 0.0882 | 0.1309 | 0.1830 | 0.4210 |
| DAX | 0.1438 | 0.2030 | 0.2916 | 0.6629 | 0.1278 | 0.1842 | 0.2724 | 0.6552 |
| Hang Seng | 0.1738 | 0.2667 | 0.4147 | 1.1749 | 0.1735 | 0.2700 | 0.4201 | 1.2130 |
| Nikkei 225 | 0.1037 | 0.1490 | 0.2095 | 0.4546 | 0.1551 | 0.1522 | 0.2049 | 0.4079 |
| **ES** | | | | | | | | |
| S&P500 | 0.0976 | 0.1598 | 0.2498 | 0.7789 | 0.1110 | 0.1742 | 0.2633 | 0.7440 |
| FTSE100 | 0.1089 | 0.1609 | 0.2406 | 0.6581 | 0.0906 | 0.1312 | 0.1914 | 0.4321 |
| DAX | 0.1462 | 0.2112 | 0.2921 | 0.6795 | 0.1335 | 0.1999 | 0.2811 | 0.6875 |
| Hang Seng | 0.2025 | 0.3069 | 0.4775 | 1.3617 | 0.2017 | 0.3197 | 0.4932 | 1.4061 |
| Nikkei 225 | 0.1036 | 0.1465 | 0.2047 | 0.4478 | 0.1071 | 0.1446 | 0.1934 | 0.3851 |

*Notes*: Estimates in daily % return terms based on a semi-parametric bootstrap with 5000 resamples using the parameter values shown in Table 1. $\alpha$ is the confidence level and the holding period is 1 day.

**Table 4: 90% Confidence Intervals for VaRs and Expected Shortfalls**

|  | $\alpha$=0.98 | $\alpha$=0.99 | $\alpha$=0.995 | $\alpha$=0.999 |
|---|---|---|---|---|
| | **VaR of long position** | | | |
| S&P500 | [0.9476 1.0560] | [0.9294 1.0769] | [0.9072 1.1025] | [0.8243 1.2253] |
| FTSE100 | [0.9384 1.0651] | [0.9252 1.0805] | [0.9082 1.1015] | [0.8413 1.2015] |
| DAX | [0.9327 1.0691] | [0.9245 1.0797] | [0.9107 1.0965] | [0.8602 1.1638] |
| Hang Seng | [0.9346 1.0731] | [0.9179 1.0859] | [0.9001 1.1140] | [0.8250 1.2214] |
| Nikkei 225 | [0.9479 1.0533] | [0.9376 1.0659] | [0.9258 1.0810] | [0.8825 1.1386] |
| | **VaR of short position** | | | |
| S&P500 | [0.9348 1.0693] | [0.9224 1.0870] | [0.9002 1.1120] | [0.8318 1.2177] |
| FTSE100 | [0.9434 1.0580] | [0.9310 1.0724] | [0.9202 1.0880] | [0.8729 1.1574] |
| DAX | [0.9351 1.0646] | [0.9271 1.0785] | [0.9106 1.0959] | [0.8557 1.1731] |
| Hang Seng | [0.9325 1.0713] | [0.9192 1.0882] | [0.8959 1.1137] | [0.8222 1.2269] |
| Nikkei 225 | [0.9440 1.0587] | [0.9385 1.0647] | [0.9263 1.0762] | [0.8935 1.1186] |
| | **ES of long position** | | | |
| S&P500 | [0.9519 1.0515] | [0.9338 1.0711] | [0.9141 1.0985] | [0.8334 1.2221] |
| FTSE100 | [0.9476 1.0536] | [0.9377 1.0671] | [0.9198 1.0903] | [0.8542 1.1776] |
| DAX | [0.9499 1.0517] | [0.9399 1.0654] | [0.9295 1.0805] | [0.8821 1.1413] |
| Hang Seng | [0.9450 1.0592] | [0.9315 1.0746] | [0.9111 1.0978] | [0.8366 1.2073] |
| Nikkei 225 | [0.9583 1.0420] | [0.9509 1.0527] | [0.9382 1.0653] | [0.8994 1.1156] |
| | **ES of short position** | | | |
| S&P500 | [0.9482 1.0543] | [0.9312 1.0731] | [0.9139 1.0944] | [0.8450 1.1996] |
| FTSE100 | [0.9559 1.0469] | [0.9456 1.0581] | [0.9313 1.0752] | [0.8878 1.1366] |
| DAX | [0.9516 1.0501] | [0.9398 1.0662] | [0.9278 1.0831] | [0.8737 1.1536] |
| Hang Seng | [0.9469 1.0580] | [0.9297 1.0777] | [0.9072 1.1011] | [0.8363 1.2112] |
| Nikkei 225 | [0.9587 1.0426] | [0.9521 1.0508] | [0.9426 1.0623] | [0.9107 1.1018] |

*Notes*: Estimates in daily % return terms based on a semi-parametric bootstrap with 5000 resamples using the parameter values shown in Table 1. $\alpha$ is the confidence level and the holding period is 1 day. Bounds of confidence intervals are standardised (i.e., divided) by the means of the boostrapped estimates.

**Table 5: Approximation Errors (%) of Numerical Integration Estimates of Exponential Spectral Risk Measure**

| Numerical integration method | Number of slices (N) | | | | |
|---|---|---|---|---|---|
| | 1000 | 10000 | 100000 | 1000000 | 10000000 |
| Trapezoidal rule | -16.38 | -2.48 | -0.34 | -0.04 | 0.00 |
| Simpson's rule | -16.67 | -2.51 | -0.34 | -0.04 | 0.00 |
| Niederreiter quasi MC | -14.27 | -3.49 | -0.56 | -0.07 | 0.00 |
| Weyl quasi MC | -14.27 | -3.49 | -0.56 | -0.07 | 0.00 |

*Notes*: Estimates are based on the mean long-position parameters in Table 1 (i.e., $\beta = 0.914$, $\xi = 0.082$, threshold =1.9, and $N_u = 249$), and $R = 100$. Errors are assessed against a 'true' value of 4.595 obtained using the trapezoidal rule with $N$ = 20 million. Estimates of pseudo MC errors are standard derivations of sample pseudo risk estimates based on samples of size 100.

**Table 6: Estimates of Exponential Spectral Risk Measures for Futures Positions**

| Futures index | $R = 20$ | $R = 100$ | $R = 200$ |
|---|---|---|---|
| **Spectral-exponential risk of long position** | | | |
| S&P500 | 2.2965 | 3.5143 | 4.156 |
| FTSE100 | 2.2871 | 3.6629 | 4.326 |
| DAX | 3.0894 | 5.0365 | 5.884 |
| Hang Seng | 3.8460 | 6.3850 | 7.651 |
| Nikkei 225 | 2.9378 | 4.3428 | 4.940 |
| **Spectral-exponential risk of short position** | | | |
| S&P500 | 2.2549 | 3.6731 | 4.380 |
| FTSE100 | 2.2973 | 3.5165 | 4.053 |
| DAX | 2.9767 | 4.7331 | 5.533 |
| Hang Seng | 3.8804 | 6.4284 | 7.713 |
| Nikkei 225 | 2.9355 | 4.4180 | 5.006 |

*Notes*: Estimates based on the parameter values shown in Table 1, using the trapezoidal integration method with $N$=1 million.

**Table 7: Standard Errors for Exponential Spectral Risk Measures**

| Futures index | Long position | | | Short position | | |
|---|---|---|---|---|---|---|
| | $R = 20$ | $R = 100$ | $R = 200$ | $R = 20$ | $R = 100$ | $R = 200$ |
| S&P500 | 0.1575 | 0.5273 | 0.8862 | 0.1662 | 0.5636 | 0.9247 |
| FTSE100 | 0.1626 | 0.5405 | 0.8960 | 0.1538 | 0.5009 | 0.7988 |
| DAX | 0.2226 | 0.7363 | 1.1901 | 0.2117 | 0.7103 | 1.1483 |
| Hang Seng | 0.2809 | 0.9724 | 1.6352 | 0.2866 | 0.9934 | 1.6845 |
| Nikkei 225 | 0.1950 | 0.6018 | 0.9576 | 0.1969 | 0.6173 | 0.9702 |

*Notes*: Estimates in daily % return terms based on a semi-parametric bootstrap with 5000 resamples using the parameter values shown in Table 1. The holding period is 1 day.

**Table 8: 90% Confidence Intervals for Exponential Spectral Risk Measures**

| Futures index | $R = 20$ | $R = 100$ | $R = 200$ |
|---|---|---|---|
| | **Long position** | | |
| S&P500 | [0.8895  1.1143] | [0.7682  1.2576] | [0.6769  1.3806] |
| FTSE100 | [0.8860  1.1221] | [0.7704  1.2505] | [0.6822  1.3594] |
| DAX | [0.8844  1.1231] | [0.7697  1.2512] | [0.6834  1.3597] |
| Hang Seng | [0.8824  1.1235] | [0.7661  1.2610] | [0.6742  1.3758] |
| Nikkei 225 | [0.8938  1.1123] | [0.7783  1.2314] | [0.7004  1.3378] |
| | **Short position** | | |
| S&P500 | [0.8825  1.1257] | [0.7622  1.2654] | [0.6768  1.3737] |
| FTSE100 | [0.8931  1.1126] | [0.7711  1.2454] | [0.6963  1.3394] |
| DAX | [0.8854  1.1167] | [0.7648  1.2554] | [0.6816  1.3556] |
| Hang Seng | [0.8818  1.1258] | [0.7623  1.2660] | [0.6602  1.3816] |
| Nikkei 225 | [0.8924  1.1117] | [0.7790  1.2424] | [0.6958  1.3277] |

*Notes*: Estimates in daily % return terms based on a semi-parametric bootstrap with 5000 resamples using the parameter values shown in Table 1. The holding period is 1 day. Bounds of confidence intervals are standardised (i.e., divided) by the means of the booststrapped estimates.

**Figure 1: Exponential Risk-Aversion Function for Various Values of the
Coefficient of Absolute Risk Aversion**

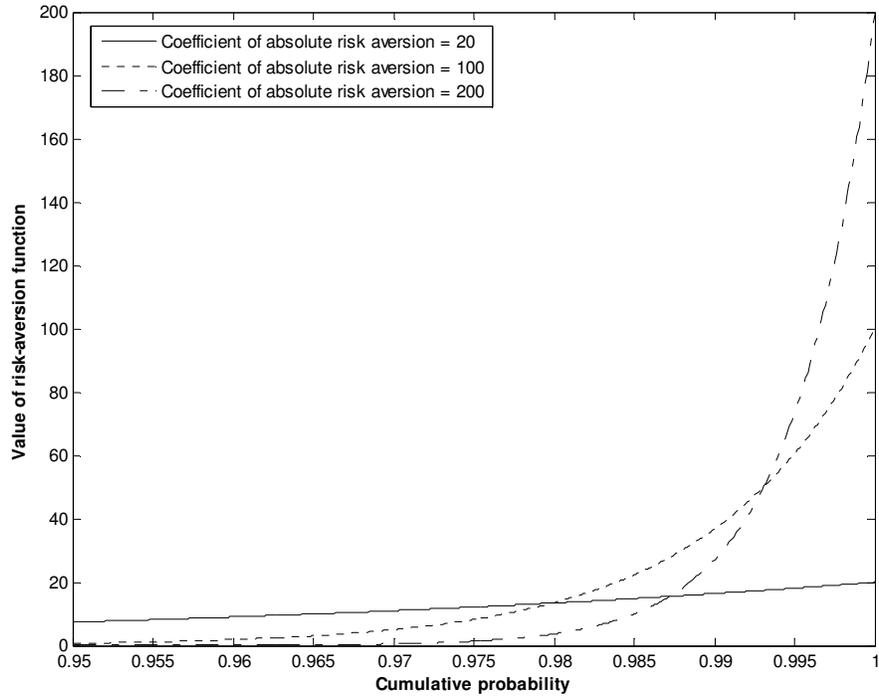

Notes: Based on equation (4) in the text.

# Figure 2: QQ Plots for Futures Return Indexes

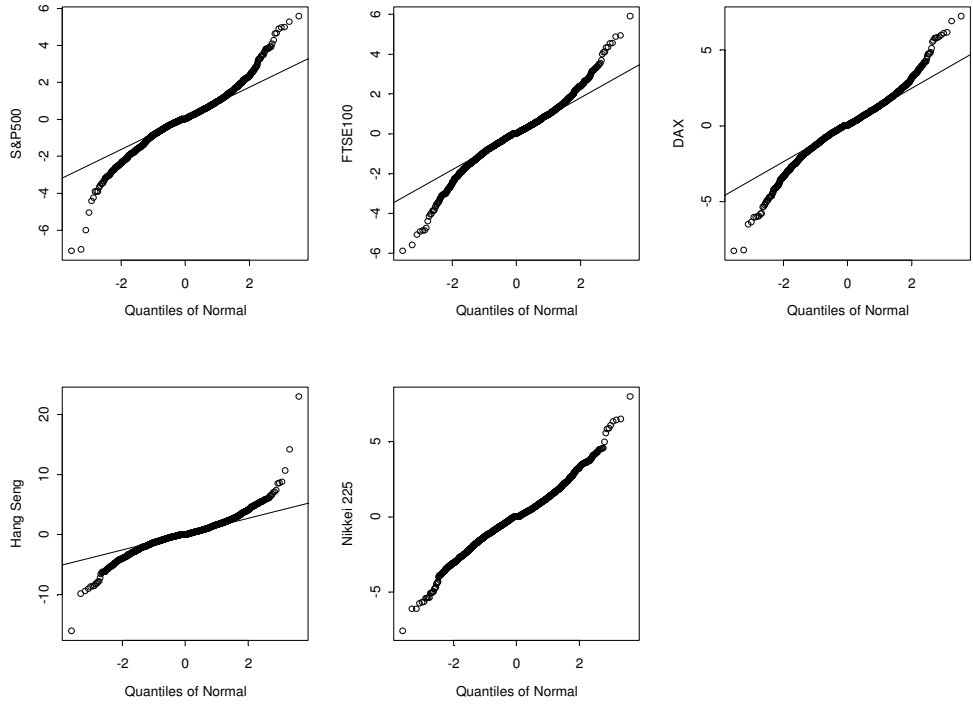

Notes: Quantiles of the respective empirical return distribution against those of normal distributions.

**Figure 3: Generalised Pareto VaRs of Futures Positions at Extreme Confidence Levels**

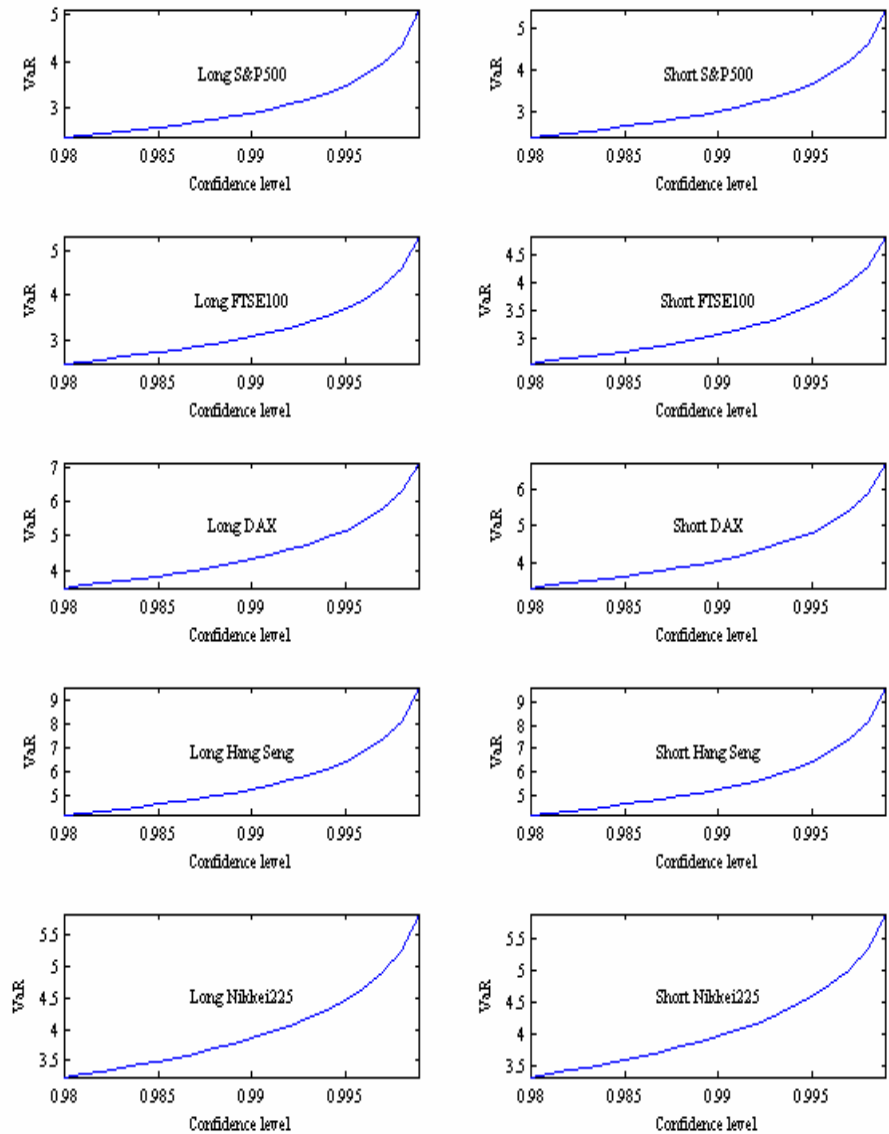

Notes: Based on the parameter values given in Table 1.

**Figure 4: Generalised Pareto Expected Shortfalls of Futures Positions at Extreme Confidence Levels**

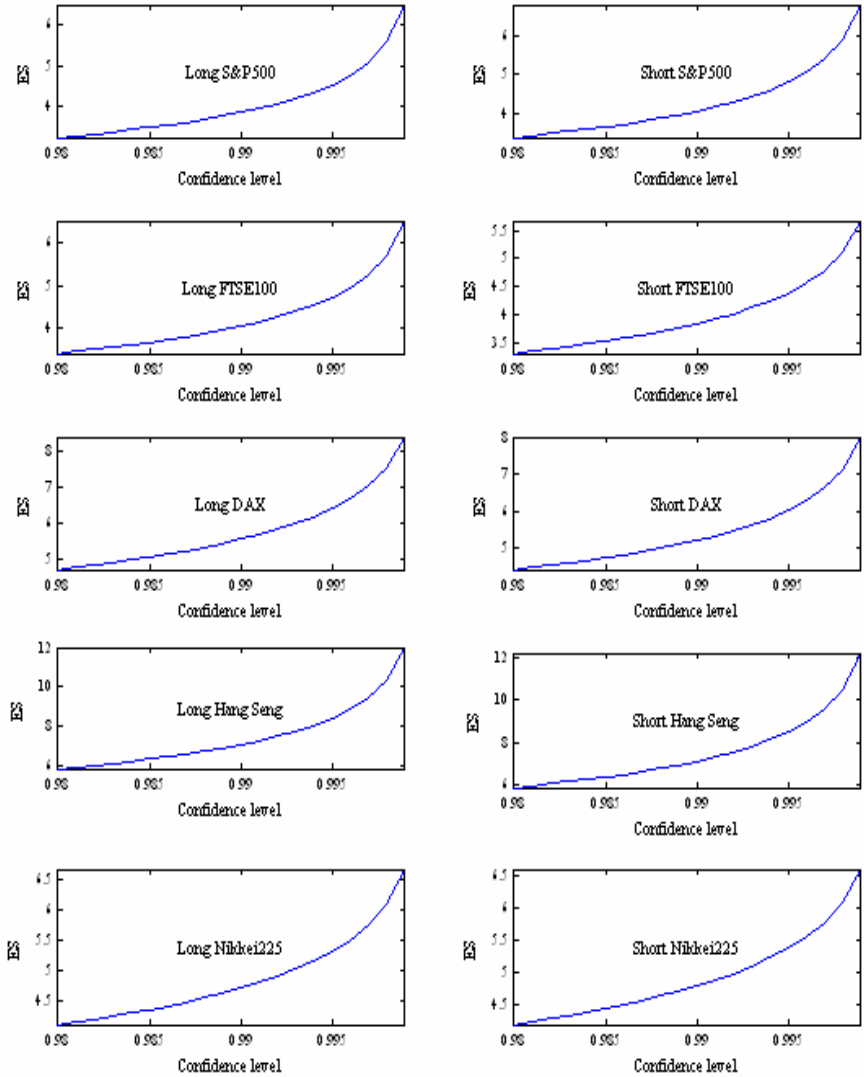

Notes: Based on the parameter values given in Table 1.

**Figure 5: Plots of Estimated Exponential Spectral Risk Measures Against the Number of Slices, *N***

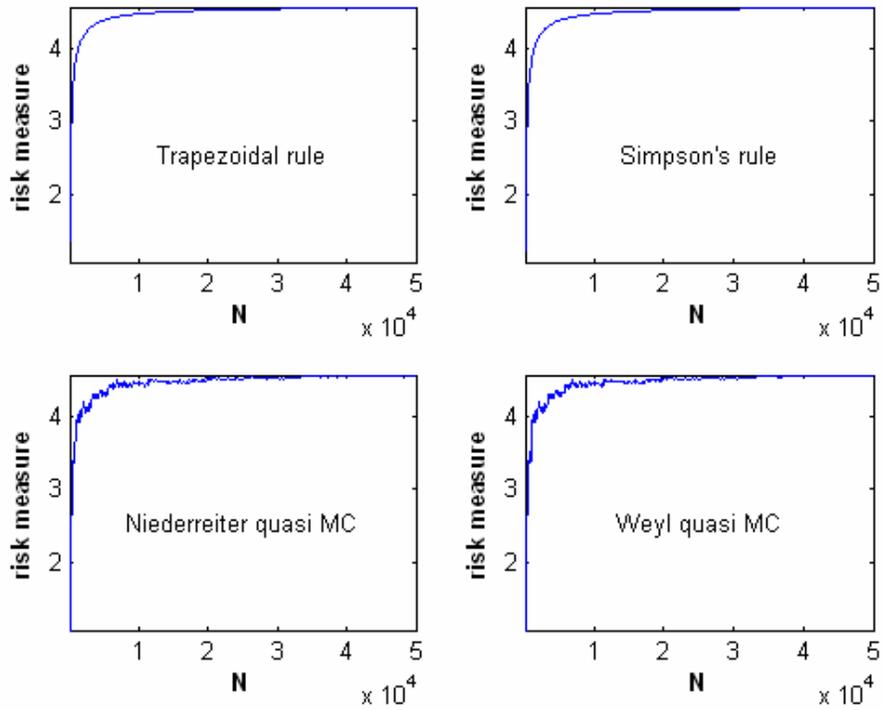

Notes: Each plot shows the estimated spectral-exponential risk measure against *N*, where *N* covers the range 100 to 50000 in steps of 100, obtained using the numerical integration routines shown on each plot. Estimates are based on the mean long-position parameters in Table 1 (i.e., $\beta = 0.914$, $\xi = 0.082$, threshold=1.9, and $N_u = 249$) along with *R = 100*.

**Figure 6:  Exponential Spectral Risk Measures of Futures Positions**

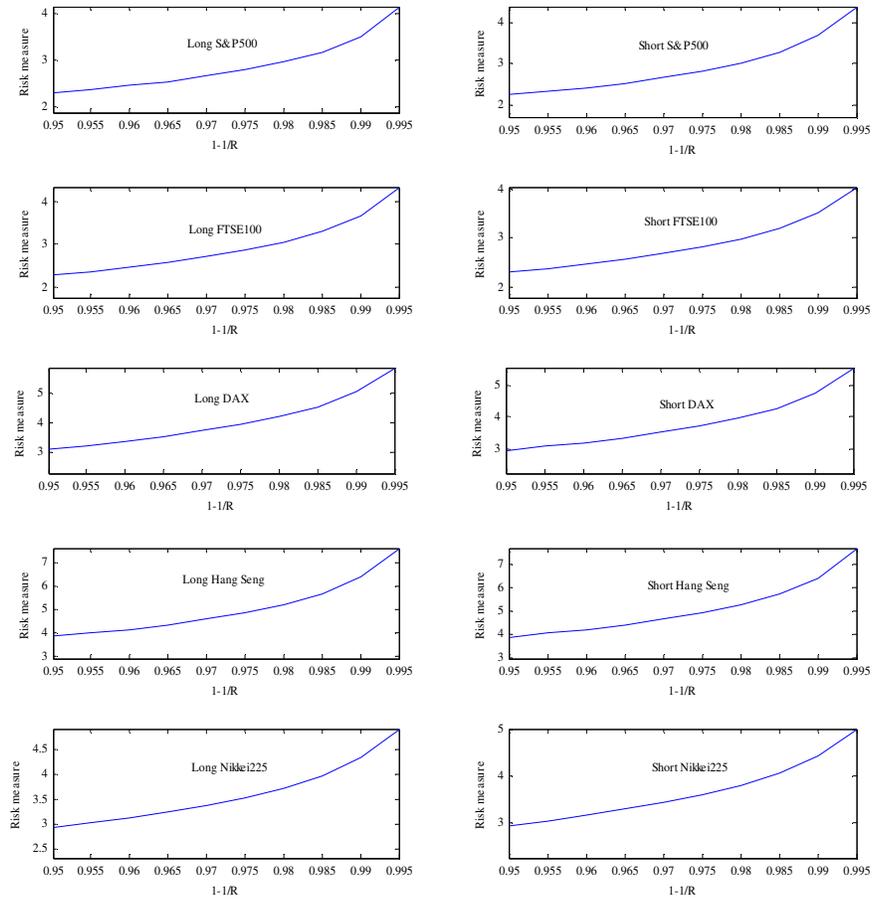

Notes: Based on the parameter values given in Table 1.